\documentclass{article}
\usepackage[a4paper]{geometry}
\usepackage[english]{babel}
\usepackage{float}
\usepackage{graphicx}
\usepackage{tabularx}
\usepackage{amsmath}
\usepackage{natbib}
\usepackage[hidelinks]{hyperref}
\usepackage{appendix}
\usepackage{rotating} 
\usepackage{adjustbox} 
\usepackage{lscape}
\usepackage{threeparttable} 
\usepackage{caption}

\title{Correcting inferences for volunteer-collected data with geospatial sampling bias}
\author{Peter Lugtig \and Annemarie Timmers \and Erik-Jan van Kesteren}
\date{Utrecht University, Department of Methodology \& Statistics\\Utrecht, Netherlands}

\begin{document}

\maketitle
\begin{abstract}
Citizen science projects in which volunteers collect data are increasingly popular due to their ability to engage the public with scientific questions. The scientific value of these data are however hampered by several biases. In this paper, we deal with geospatial sampling bias by enriching the volunteer-collected data with geographical covariates, and then using regression-based models to correct for bias. We show that night sky brightness estimates change substantially after correction, and that the corrected inferences better represent an external satellite-derived measure of skyglow. We conclude that geospatial bias correction can greatly increase the scientific value of citizen science projects.\\

\end{abstract}

\section{Introduction}
Citizen science projects in which volunteers collect data in scientific projects are increasingly popular due to their ability to engage the public with a scientific question. Moreover, scientists benefit from engaging volunteers by learning from local knowledge, and gaining access to data that would otherwise be very expensive or impossible to collect. In the Netherlands alone, more than 130 citizen science projects are currently ongoing or recurring \citep{timmers_annemarie_2021_4724570}. The value of these projects in terms of outreach is clear: projects such as the national bird count are widely known and bring attention to the field, in this case ecology \citep{land2016citizen,van2016citizen}.

Many citizen science projects have an inherent geographic component, as citizens have knowledge and are interested in their immediate environment: citizen science projects focus for example on citizens taking pictures of plastic \citep{rambonnet2019making}, monitoring earthquakes \citep{finazzi2020fulfilling}, documenting natural life and ecology \citep{mckinley2017citizen,brown2019potential,petersen2021species}, or looking at the brightness of the night sky \citep{birriel2014analysis} in their neighborhood. Many citizen science projects stem from local initiatives, yet there are also many projects that run at a national, or even international scale \citep{birriel2014analysis,newman2012future}.

While the value of these projects in terms of outreach is clear, their use as a scientific instrument is more controversial \citep{bonney2014next}. For some projects, it has been possible to use citizen scientists to predict earthquake risk in real time, allowing policy makers to design faster or better warning systems \citep{finazzi2020fulfilling}. Other projects that explored biodiversity and pollution are being used by the United Nations for humanitarian activities and monitoring of the 2030 Sustainable Development Goals \citep{crimmins2022large,de2021critical}. When it comes to projects that focus on documentation, such as counting natural life or plastic pollution, citizen science projects could be used to better understand the number of birds or extent of plastic pollution, and their geographic diversity. 

However, in terms of scientific value citizen science projects face several difficulties. First, citizen scientists may often need training to take measurements of high quality. And even with training, professional scientists are often used to assess the quality of measurements \citep{swanson2016generalized}. Over time, citizens may become more or less active, leading to large variation over time in estimates derived from citizen science projects \citep{bowler2022temporal}. Another prominent bias -- and the focus of this paper -- is geospatial sampling bias: observations are often taken in the direct environment of citizen scientists, leading to many observations in some areas, and very sparse (or no) coverage of other areas \citep{petersen2021species}. 
Citizen science projects are often aiming to reduce geospatial sampling biases by encouraging citizen scientists to take observations in particular areas \citep{budhathoki2013motivation}. Indeed, previous articles on the future of citizen science data have focused on emerging technologies such as apps and online gaming to advance participation \citep{newman2012future}. For projects measuring air pollution in urban areas for example, projects have worked with mobile sensors \citep{adams2020spatial} installed on bikes or cars to better capture local and temporal variations \citep{sonnenschein2022agent}.
While such efforts may improve the coverage of volunteer data collection efforts in the region of interest, the problem of geospatial sampling bias in such projects remains substantial. In many citizen science projects areas of interest are often missing completely, and researchers have little or no control over where observations are taken.

In this paper, we develop and implement a generic method to deal with geospatial sampling bias by using a land-use regression model with spatial covariates. Covariates should be available at a local-level, measured without much error, be related to both the spatial selectivity, and outcome variable, and be available for the entire geographical area of inference. Our method allows citizen science projects to infer and predict to a relevant geographical unit (e.g. a province or country) even when there are limited observations in certain areas of the geographical unit. Whereas land-use regression is common in modeling spatial data in general, it has to our knowledge not been used to model sampling bias in volunteer-collected data specifically. 

This paper is structured as follows. In the next section, we first describe the Globe at Night citizen science project, which we will use as an example to illustrate our approach. Then, we describe our land-use regression model, and external data sources that we are using to correct for sampling bias. We then test different types of land-use models to the Globe at Night data and compare them to an external benchmark dataset that uses satellite data to monitor sky brightness. We finish with a conclusion and discussion of our findings. 

\section{Methods}
In order to explain our general method, we will in this section first introduce both the volunteer-collected dataset that we use as a motivational example, and two geospatial data sources that contain the covariates used for bias correction. After this we will explain how we integrate these data sources, and show the model that we propose to correct for spatial bias in volunteer-collected data.

\subsection{Data description}
\subsubsection{Globe at Night project}
\label{sec:gandata}
Globe at Night (GAN) is a citizen-science campaign started in 2006 by the National Optical Astronomy Observatory (NOAO) with the aim to raise awareness to the effects of light pollution. The campaign collects naked eye limiting magnitude (NELM) estimates of citizens that are asked to go outside an hour after the sun has set and report on the brightness of the sky. They do this by linking a monthly changing constellation to one of eight stellar maps that gradually show fainter stars. On top of the sky-brightness, citizens report on the sky conditions at the time of observation, and the time, data and location of the observation. Lastly, there is an option to enter data of a Sky Quality Meter, which allows for a more precise estimation of the brightness of the sky \citep{birriel2014analysis}. Data can either be entered into an app or on the \href{https://www.globeatnight.org/webapp/}{GAN webapp}. In 2021, the campaign received 25,551 observations from 90 countries. 

\begin{figure}[H]
    \centering
    \includegraphics[width=\textwidth]{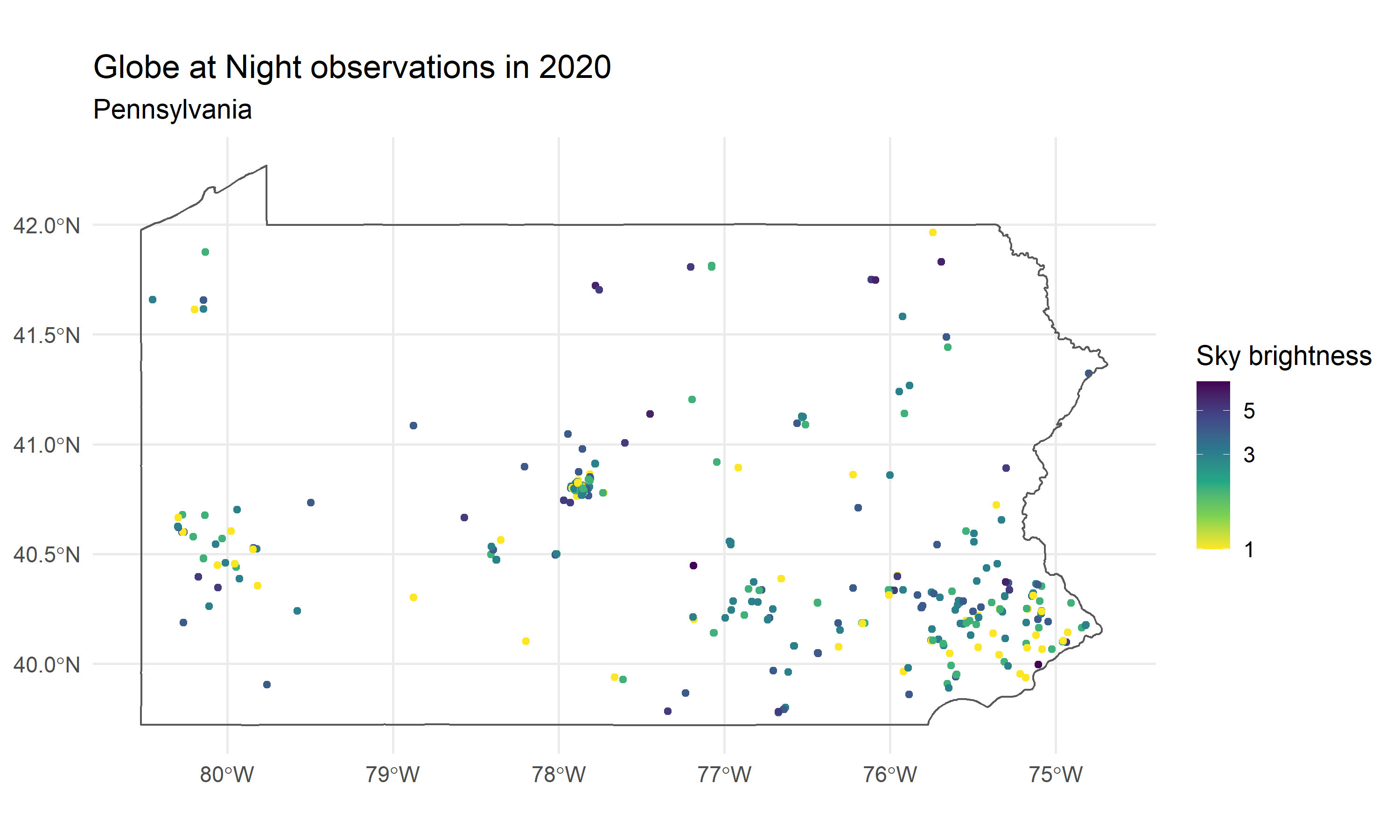}
    \caption{Locations and sky measures of the Globe at Night project in 2020 in Pennsylvania. Note that the brightness scale is reversed: higher values mean a darker sky.}
    \label{fig:raw_gan}
\end{figure}

Data is publicly available and can be viewed or downloaded by year from the \href{https://www.globeatnight.org/maps.php}{GAN website}. In order to show how citizen science data from the Globe at Night project can be used for inference, we will restrict our analysis to observations taken in the state of Pennsylvania in the United States in the year 2020. This results in 1366 observations of sky brightness, depicted in Figure \ref{fig:raw_gan}. Volunteers are asked to observe the number of stars visible around Arcturus, the brightest star in the Boötes constellation. An observation of 0 means that just Arcturus is visible, while an observation of 7 implies the sky is fully dark, and many stars are visible. Volunteers can use Magnitude Charts to assist them in making ratings.

The map in Figure \ref{fig:raw_gan} shows that observations in 2020 were not uniformly distributed across the state of Pennsylvania. Large swathes of the state are not covered by observations in the Globe at Night project, while other areas are well-covered. This is typical of citizen science or volunteer-collected data, as citizens are more likely to make observations in areas that are of special interest to them. To correct for biases in this spatial distribution, we will use external datasets as a source of geospatial covariates. In the following sections, we describe these additional datasets.

\subsubsection{Land-use data}
\label{sec:land-use data}
A main source of geospatial covariates in geostatistics is so-called ``land-use data''. For the region of interest in the motivating example in this paper, land-use data is available from the Multi-Resolution Land Characteristics Consortium \citep[MRLC;][]{wickham2014multi}. This consortium publishes the United States National Land Cover Database \citep{yang2018new}, either for direct download or online via an Open Geospatial Consortium (OGC) Web Map Service (WMS).

In summary, the land cover database contains high-resolution images where each pixel is assigned to a specific land cover type such as \emph{forest}, \emph{developed}, or \emph{water} (see Table \ref{tab:landuse_info} in Appendix \ref{app:landcover} for all labels and their descriptions). This land-cover type is the result of a model which integrates multi-spectral, multi-temporal, and spatial information from remote sensing data, i.e., aerial photography or satellite imagery. These land use data are commonly used as predictors in a \emph{land use regression model} in geospatial studies, such as for air pollution estimation \citep[e.g.,][]{hoek2008review}. 

\begin{figure}[H]
    \centering
    \includegraphics[width=\textwidth]{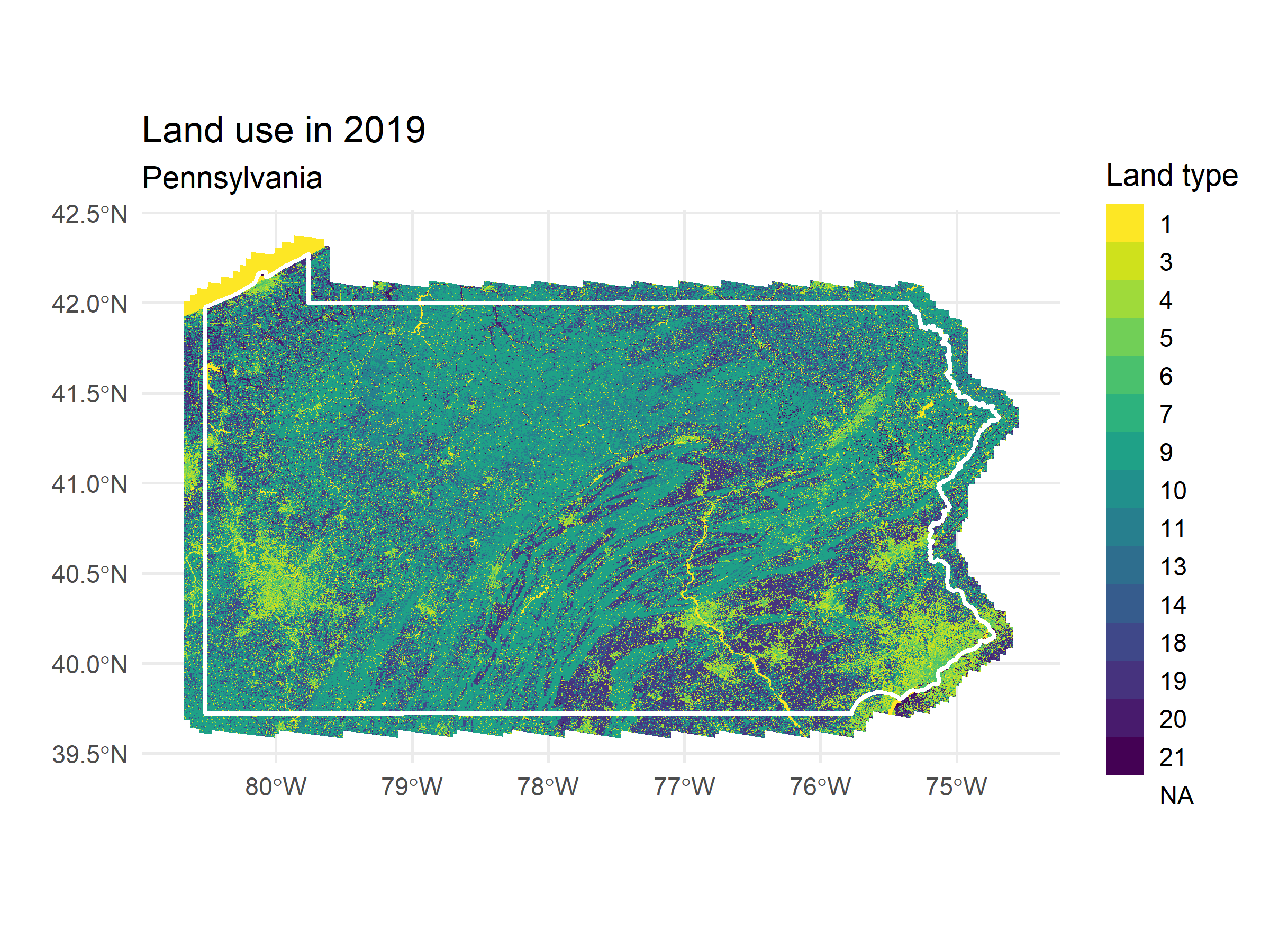}
    \caption{The 2019 version of land-use data for the general Pennsylvania area. Data are obtained from the Multi-Resolution Land Characteristics Consortium (MRLC).}
    \label{fig:raw_landuse}
\end{figure}

Figure \ref{fig:raw_landuse} shows the land use for the state of Pennsylvania and nearby locations. Land-use models operate on a raster (i.e. pixel-by-pixel) basis, assigning only a single category to a potentially diverse area. One downside of this approach is that the land-use variables may miss information crucial to light pollution and sky brightness, such as road infrastructure. Hence, we make use of another, vector-based geospatial data source to add information about the road network in Pennsylvania.


\subsubsection{OpenStreetMap}
\label{sec:osmdata}
The second source of geospatial covariates we use is OpenStreetMap (OSM). The OSM project was founded in 2004 by Steve Coast and based on the belief that a map of the world could be crowdsourced by people with local knowledge \citep{fonte2017generating}. OSM provides a free map of the world that is generated and edited by volunteers. There are three core types of data or objects in OSM: nodes, ways and relations. Nodes are geographical points in space defined by latitude, longitude and node id \citep{mooney2017review}. Ways consist of polylines and polygons, which are composed out of at least two or three nodes, respectively. Each of these three objects are described by one or more attributes or tags \citep{mooney2017review}. Tags are defined by the users themselves and decided on democratically. Each tag is accompanied by a wiki where the use of the tag is outlined, as well as some assumptions regarding the use of the tag. For example, the highway wiki mentions that what is considered a primary road may vary depending on the region of the world. 
Data are available under the Open Database License can be downloaded from the \href{https://planet.openstreetmap.org/}{OSM website}. We extracted data of the tag \texttt{highway=motorway} for the state of Pennsylvania to use as a geospatial covariate. The \href{https://wiki.openstreetmap.org/wiki/Key:highway}{highway wiki} defines a motorway as \textit{"A restricted access major divided highway, normally with 2 or more running lanes plus emergency hard shoulder. Equivalent to the Freeway, Autobahn, etc..."}. Starting in 2009, highway tags are defined by their importance to the road grid, not by physical attributes \citep{Openstreetmap}. Out of the 76 highway tags, the motorway is the most important one. In addition to importance, the motorway tag implies road quality as well as intended usage. Lights around motorways not only directly affect light pollution; because motorways are concentrated in urban areas, they can also serve as more general proxy information for areas with high concentrations of buildings \citep{kuechly2012aerial}.

\begin{figure}[H]
    \centering
    \includegraphics[width=\textwidth]{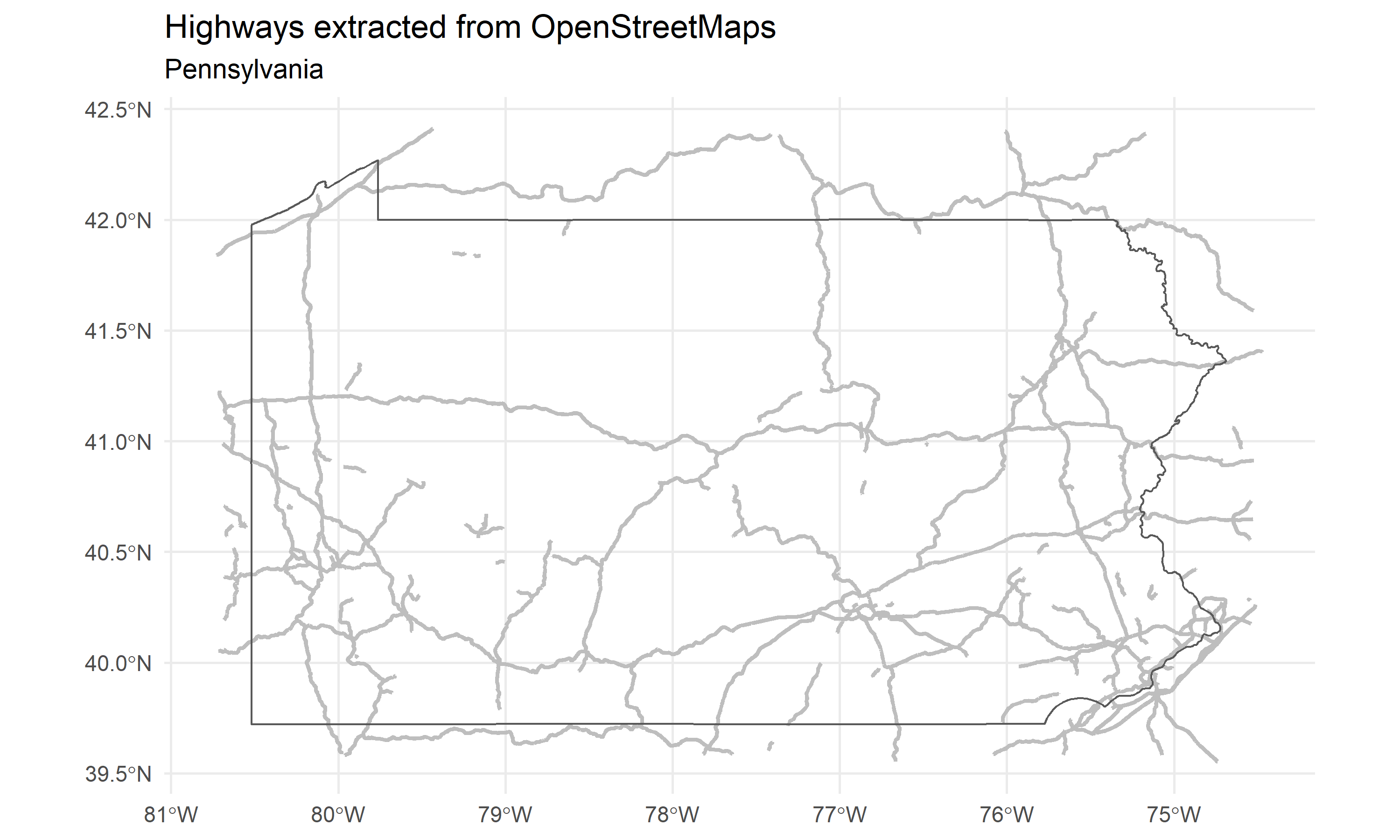}
    \caption{Polylines of OpenStreetMap-derived highways (of type \texttt{motorway}) in Pennsylvania. Highways are a source of light pollution, even in nature-abundant areas which are otherwise dark.}
    \label{fig:raw_hwy}
\end{figure}

\subsection{Geospatial enrichment procedure}
As a first step in our analysis, we enrich the Globe at Night data with data from both OpenStreetMap and land use sources. First, we used \texttt{osmenrich} \citep{van_kesteren_erik_jan_2021_4534188} to compute a (weighted) number of motorways in the neighbourhood of each observation (10 kilometer radius). Using the standard Gaussian kernel implemented in \texttt{osmenrich}, higher weight is given to motorways close to the observation. For land use we compute the proportion of each land-use type in a circular area of 25 square kilometers around each observation (see Figure \ref{fig:landuse_GaN} in the Appendix).

Next, we divide up the state of Pennsylvania into a regular \emph{prediction grid} of 4843 5x5 km cells. On this prediction grid, we then also perform the same enrichment procedure, resulting in our OSM covariates and land-use covariates computed for each cell in the grid. Figure \ref{fig:landuse_state} in the Appendix shows the grid cells overlayed on the land-use data.

\subsection{Land-use regression to account for sampling bias}
For geographic data, land-use regression models are typically used to infer from specific observations to entire geographical areas \citep{hoek2008review}. One area where land-use regression models have been used frequently is in estimating air pollution levels. Air pollution sensors are located in specific urban locations, and because of the fact that air pollution levels may differ a lot depending on very local circumstances, as well as over time, interpolation methods such as ordinary kriging do not suffice to study air pollution across an entire area, nor to predict in what specific areas air pollution is likely to exceed a threshold. 

Land-use regression models typically model the geographic area of interest as a raster. Characteristics such as population density or the presence of big roads or industrial areas at each raster cell can then serve as covariates to predict the variable of interest for those areas of the raster where no observations are made \citep{hoek2008review}. Residuals in the regression model are in practice likely to be correlated between areas that are in close proximity. In order to deal with this land-use models can incorporate kriging methods \citep{wu2018hybrid} that allow residuals between areas in close proximity to be correlated. Residuals can either be caused by very local deviations from the spatial pattern (e.g. by a local high-emitting source of pollution), or be caused by measurement error. Kriging ensures that local deviations not accounted for by the land-use regression model are incorporated in the model.

\subsection{Analysis model for the Globe at Night data}
After enriching each 5x5 area in Pennsylvania with data about land-use and infrastructure from OSM, we create regression models with four different levels of covariate information:


\begin{enumerate}
    \item A null model without covariates. In this model, we take the mean sky brightness in every area where Globe at Night observations were made, and take the overall mean for areas not covered by Globe at Night. This model serves as a baseline model against which to compare the following models.
    \item A model using United States National Land Cover Database covariates. In this model, the land-use proportions in for each observation are added as predictors. 
    \item A model using OpenStreetMaps road data. In this model, the number of motorways in a 10km radius is added as a predictor.
    \item A combined model using the predictors from both sources. 
\end{enumerate}

In addition to these covariates, we also correct for cloud cover, a categorical variable included in the Globe at Night data; and moon illumination, computed from the location and time information associated with each observation using the R package \texttt{suncalc} \citep{thieurmel2022suncalc}. We also run each model with and without kriging to account for autocorrelations in observations that are located closely together in space. We use universal kriging with a Matérn covariance kernel estimated on the (conditional) residuals, as implemented in the package \emph{gstat} \citep{graler2016spatio}. 



We validate the eight resulting models in two distinct ways. First, we assess the prediction performance of each model using mean square error (MSE), computed through leave-one-out cross-validation \citep[LOOCV,][]{giraldo2011ordinary}. We call this \textit{internal validation}. Second, we additionally validate the models using an external data source. \textit{External validation data} for citizen science datasets are often not available, but for the Globe at Night data, we can use satellite imagery data to benchmark our model performance. We do this by examining both the correlation of the model estimates with the external data, and the variance in satellite-derived sky brightness that is explained by our models as measured by $R^2$. This external validation step is crucial for assessing the real-world impact of the bias-corrective effects of our proposed data enrichment procedure.

Finally, we evaluate the effect of our geospatial sampling bias corrections on inference. The quantity of interest here is the overall average sky brightness in the state of Pennsylvania in the year 2021. We calculate the mean sky brightness before and after correcting, along with measures of uncertainty, to show how the estimates of this quantity may change.

All our analyses were conducted in R 4.1.2 \citep{R}. Data and code to replicate our procedures are available on: \url{https://doi.org/10.5281/zenodo.6597859} \citep{erik_jan_van_kesteren_2022_6597859}.

\section{Results}

Table \ref{tab:results} shows the results of several models predicting sky brightness for each 5x5 km area in the state of Pennsylvania, using the Globe at Night data as input data, and varying inclusion of geospatial covariates.
The mean model takes the mean of sky brightness, and serves as a baseline model against which to compare our other models. 

\subsection{Internal validation}
\label{sec:internalvalid}

The MSE's for internal validation  in Table \ref{tab:results} show that the model that combines all spatial covariates and uses kriging performs the best (MSE = 0.956), although the differences between the individual models are small. Additionally, The MSE is lower for all models with kriging, compared to their counterpart without kriging. Moreover, the MSE is lower for the models that only included the land-use covariates, compared to the models that added the OSM motorways as a covariate.

\begin{table}[H]
\caption{\label{tab:results} Model comparison using an internal metric (Leave-one-out cross-validated mean square error) and an external metric (Spearman rank correlation with satellite-derived log-skyglow values).}
\renewcommand{\arraystretch}{1.2}
\centering
\begin{tabular}{rllccc}
  \hline
  & & & Internal validation & \multicolumn{2}{c}{External validation} \\
  \cline{4-6}
  Model & Covariates & Kriging & LOOCV MSE & Spearman's $\rho$ & $R^2$ \\ 
  \hline
  1 & Mean     & No  & -              & 0.031          & 0.007 \\ 
  2 & Mean     & Yes & 1.056          & 0.186          & 0.066 \\ 
  3 & Land-use & No  & 1.067          & 0.526          & 0.183 \\ 
  4 & Land-use & Yes & 0.994          & 0.508          & 0.217 \\ 
  5 & OSM      & No  & 1.203          & \textbf{0.642} & \textbf{0.353} \\ 
  6 & OSM      & Yes & 1.033          & 0.438          & 0.302 \\ 
  7 & Combined & No  & 1.064          & 0.548          & 0.217 \\ 
  8 & Combined & Yes & \textbf{0.956} & 0.474          & 0.182 \\ 
  \hline
\end{tabular}
\end{table}

\begin{figure}
 \centering  \includegraphics[width=0.9\textwidth]{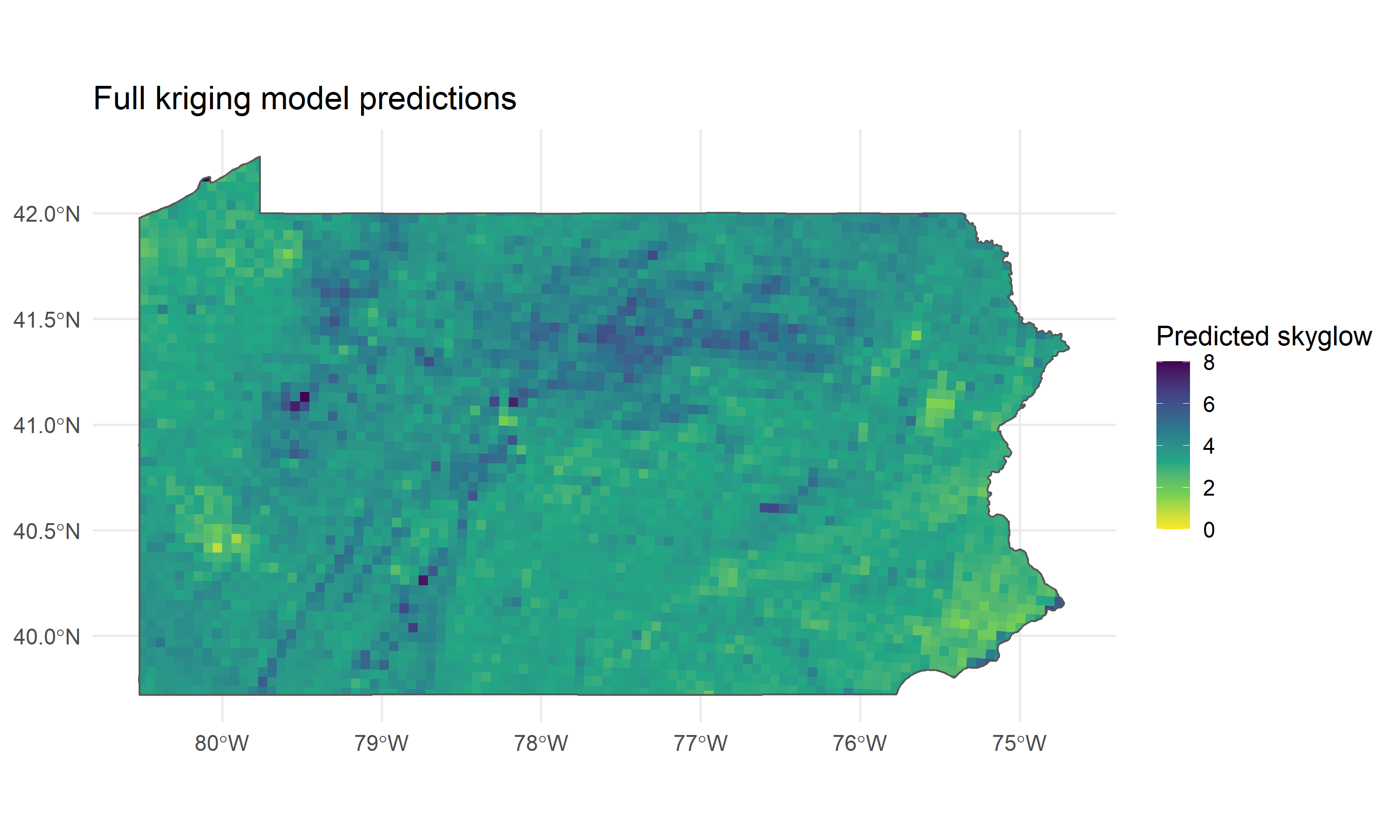}
    \caption{Predicted sky brightness in Pennsylvania using a model with land-use and OSM covariates, with kriging.}
    \label{fig:predict}
\end{figure}

Figure \ref{fig:predict} shows the predicted sky brightness  of the best performing model (model 8) mapped onto the 5x5 grid cells of the state of Pennsylvania. Brighter colours reflect brighter nightskies in an area. 

\subsection{External validation}
The Globe at Night project allows for external validation of our inferences in addition to the internal validation discussed in Section \ref{sec:internalvalid}. For this, we have downloaded and processed satellite imagery data of light radiance in Pennsylvania in 2020. Specifically, we use median nighttime cloud-free light intensity from 2020 (see Appendix \ref{app:radiance}), which was collected by the Visible Infrared Imaging Radiometer Suite (VIIRS) on the Suomi National Polar-Orbiting Partnership (Suomi NPP) spacecraft \citep{elvidge2021annual}. This high-resolution radiance map is converted into an expected average skyglow for each cell in our 5x5 grid by using ``Walker's law'' \citep{walker1977effects}: skyglow is inversely proportional to the distance of the light source (in km) to the power 2.5 \citep[see also][Fig. 10]{duriscoe2007measuring}.

\begin{figure}[H]
    \centering
    \includegraphics[width=\textwidth]{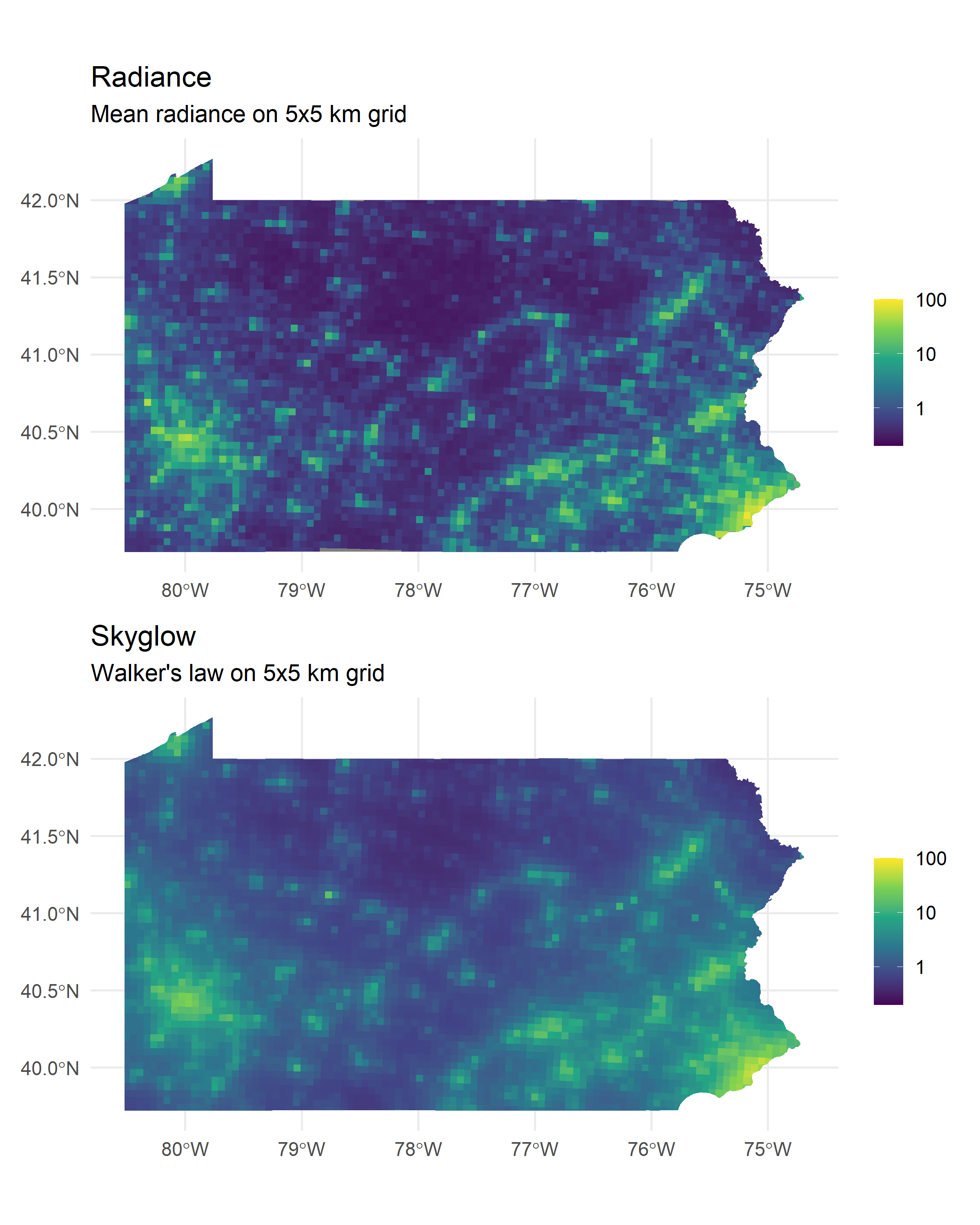}
    \caption{Satellite-derived Radiance (upper panel) and skyglow (lower panel) at our chosen grid cells in Pennsylvania. Skyglow values are processed from nighttime satellite radiance data using an inverse distance relationship.}
    \label{fig:skyglow}
\end{figure}

The resulting satellite-derived radiance- and skyglow levels are shown in Figure \ref{fig:skyglow}. Transforming the radiance  transformation captures the fact that light from a city scatters through the atmosphere to reach otherwise dark areas of the state. In the image, this is shown as a smoothing or blurring of light-intensive areas such as cities and towns.

The satellite-derived skyglow is measured in terms of the physical property of light radiance, whereas the Globe at Night sky brightness is measured using volunteer perceptions. To compare these our model predictions to the satellite derived data of the same underlying phenomenon, we use the Spearman rank correlation. These rank correlations are shown in the final column of Table \ref{tab:results}.

For both external validation measures, the mean models perform the worst as they have the smallest correlation to the radiance map and $R^2$. Kriging, however, does improve the performance of the mean model as it increases the correlation up from $\rho_s =$ 0.031 to $\rho_s =$ 0.186 and $R^2$ from 0.007 to 0.066. Next, the land-use models perform similarly and kriging does not seem to make a difference in the correlation of the estimates to the radiance map, nor was there a large difference in $R^2$ values. This contrasts the findings of the OSM models as kriging results in a correlation of $\rho_s =$ 0.438, 0.204 $\rho_s$ smaller than the model without kriging. With a difference of $R^2 =$ 0.051, the explained variance is affected to a smaller extent. We observe the same for the combined model, though  not the same extent. Overall, both the correlation to the external data and the proportion of explained variance are the strongest and largest for the OSM model without kriging: $\rho_s = $ 0.642 and $R^2 =$ 0.353. This is remarkable, as this model performed the worst in terms of internal validation. 

\subsection{Inference}

 We can now use both the observed  sky brightness statistics from the small set of observations made by citizen scientists, and the predicted sky brightness along with prediction uncertainty from model 8 to predict the mean sky brightness for the state of Pennsylvania, and evaluate the effect of our sampling bias corrections. For this we use the model with the best internal performance (model 8), as it is not common to have external validation data available for our method
 
Using the observations alone, we find that the mean sky brightness is 2.48 ± 0.034 (s.e.). 
After inference to the entire state of Pennsylvania using land-use regression, we find that sky brightness is on average more than 1 point higher at 3.67 ± .019 (s.e.). The standard error for the inferred model is smaller as many predicted observations are relatively close to the state-level mean, leading to a between-grid cell standard error of the mean of .005, and a within grid-cell standard error reflecting prediction uncertainty of .014.

\section{Discussion}
This paper shows that geospatial selection bias that is often present in citizen science data can be mapped and corrected for using publicly available data sources such as OpenStreetMap and land-use data. 

We use the Globe at Night dataset to show that a very selective set of observations about the night sky brightness from the state of Pennsylvania can be used to map the sky brightness for the entire state with a high degree of accuracy. Our method is generic and can be used in many other applications of citizen science datasets where observations are geospatial. The application can a be adapted to many kinds of citizen science topics. For example, our methods allows for inference from citizen science projects around bird counts to entire countries, but also to specific types of landscapes or regions. Another application would be documenting  plastic pollution in the environment. Here a selective set of observations in some cities can be generalized to urban environments in general.

We see two main ways in which our method can be used to improve citizen science projects for policy making. First, it allows inferences to be made. For our application, we can estimate the average sky brightness for the state of Pennsylvania for example. For projects in ecology, it would allow scientists to estimate population totals. Second, and perhaps even more useful, our method can be used to make predictions. For example, it can predict in which areas of the state of Pennsylvania the night sky is darkest. Or in the context of ecology it can be used to predict areas where population counts are predicted to be particularly high or low. In the practice of citizen science, predictions can also be used to identify areas where there is a lot of uncertainty about the prediction. A next step would then be to encourage citizen scientists to take additional observations in areas with high uncertainty. This would then in turn improve the inference, and predictions from the model.

Despite the promises of the method we propose for inference for citizen science datasets, there are several limitations to our method. First, our method relies on citizen scientists documenting accurately what they observe. In our application we have assumed that the measurements of sky brightness contain no measurement error, even though there is evidence that observations taken at the same location by different people, or at different days of the year contain sometimes large variability. We have in our application taken measurements at face-value, which in practice means that averages are taken. There are methods to account for and possibly correct for such measurement errors as well, but in our dataset, the number of observations is not large enough to formally estimate or account for such measurement error.
Second, our method relies on the fact that measurements are all documented. Citizen science projects often encourage observers to record everything they observe, for example by specifying on what time and what location observations should be taken. However, we cannot exclude the possibility that in our application not all measurements have been recorded. It is possible for example that citizen scientists only record the sky brightness when they find the night sky to be particularly bright or dark. Similarly, in projects where for example pictures are taken of plastic pollution, scientists may only take a picture when they see plastic. In such cases, our method would need to be extended to also include selectivity that is dependent on the value of the observed statistic.

Further research should also investigate what geospatial data sources can further be used to link, model and correct for selectivity in citizen science datasets. For our application, we use covariates about land use and OpenstreetMap because the built environment should be predictive of the night sky brightness. For other applications, such as ecology, geospatial covariates such as forest cover or type of water source in the area are probably more informative datasources. As long as both the citizen science data are properly geolocated, linking citizen science data to geospatial covariates from other databases should be relatively straightforward. 
The choice for which data source to use, and how to model the dataset is not straightforward however. In our application, we tested several different models using OSM and/or landuse covariates, with and without kriging to select the best model. Internal validation of the model shows that the model with all covariates and kriging performs best, whereas external validation showed a model using no kriging and just motorways as covariates from OSM worked best. But this finding is specific to our application, and may partly depend on such things as the size of the grid cells used. As our final model we yse the model that performed best using internal validation, mainly because external validation data are rarely available. 
Each future application will require a slightly different statistical model, depending on the type and number of geospatial covariates available, and the specific topic of the study. A particular risk is overfitting by using a large set of geospatial covariates in an exploratory way. We therefore recommend to carefully consider covariates used in the models, and to routinely conduct sensitivity analyses.

Citizen science holds great potential for involving the general public in science and using the power of local knowledge to learn about our local environment and society. Statistical methods hold great potential to expand the use of citizen science data to inference and policy-making. Our hope is that this paper may contribute to a further expansion of citizen science for understanding and improving our local environment.

\newpage
\bibliographystyle{rss_style/rss}
\bibliography{references}

\begin{thebibliography}{35}
\expandafter\ifx\csname natexlab\endcsname\relax\def\natexlab#1{#1}\fi
\expandafter\ifx\csname url\endcsname\relax
  \def\url#1{\texttt{#1}}\fi
\expandafter\ifx\csname urlprefix\endcsname\relax\def\urlprefix{URL: }\fi

\bibitem[{Adams et~al.(2020)Adams, Massey, Chastko and
  Cupini}]{adams2020spatial}
Adams, M.~D., Massey, F., Chastko, K. and Cupini, C. (2020) Spatial modelling
  of particulate matter air pollution sensor measurements collected by
  community scientists while cycling, land use regression with spatial
  cross-validation, and applications of machine learning for data correction.
\newblock \textit{Atmospheric Environment}, \textbf{230}, 117479.

\bibitem[{Birriel et~al.(2014)Birriel, Walker and
  Thornsberry}]{birriel2014analysis}
Birriel, J.~J., Walker, C.~E. and Thornsberry, C.~R. (2014) {Analysis of seven
  years of Globe At Night data}.
\newblock \textit{Journal of the American Association of Variable Star
  Observers}, \textbf{42}, 219--228.

\bibitem[{Bonney et~al.(2014)Bonney, Shirk, Phillips, Wiggins, Ballard,
  Miller-Rushing and Parrish}]{bonney2014next}
Bonney, R., Shirk, J.~L., Phillips, T.~B., Wiggins, A., Ballard, H.~L.,
  Miller-Rushing, A.~J. and Parrish, J.~K. (2014) Next steps for citizen
  science.
\newblock \textit{Science}, \textbf{343}, 1436--1437.

\bibitem[{Bowler et~al.(2022)Bowler, Callaghan, Bhandari, Henle,
  Benjamin~Barth, Koppitz, Klenke, Winter, Jansen, Bruelheide
  et~al.}]{bowler2022temporal}
Bowler, D.~E., Callaghan, C.~T., Bhandari, N., Henle, K., Benjamin~Barth, M.,
  Koppitz, C., Klenke, R., Winter, M., Jansen, F., Bruelheide, H. et~al. (2022)
  Temporal trends in the spatial bias of species occurrence records.
\newblock \textit{Ecography}, \textbf{2022}, e06219.

\bibitem[{Brown and Williams(2019)}]{brown2019potential}
Brown, E.~D. and Williams, B.~K. (2019) The potential for citizen science to
  produce reliable and useful information in ecology.
\newblock \textit{Conservation Biology}, \textbf{33}, 561--569.

\bibitem[{Budhathoki and Haythornthwaite(2013)}]{budhathoki2013motivation}
Budhathoki, N.~R. and Haythornthwaite, C. (2013) {Motivation for open
  collaboration: Crowd and community models and the case of OpenStreetMap}.
\newblock \textit{American Behavioral Scientist}, \textbf{57}, 548--575.

\bibitem[{Crimmins and Crimmins(2022)}]{crimmins2022large}
Crimmins, T.~M. and Crimmins, M.~A. (2022) Large-scale citizen science programs
  can support ecological and climate change assessments.
\newblock \textit{Environmental Research Letters}, \textbf{17}, 065011.

\bibitem[{Duriscoe et~al.(2007)Duriscoe, Luginbuhl and
  Moore}]{duriscoe2007measuring}
Duriscoe, D.~M., Luginbuhl, C.~B. and Moore, C.~A. (2007) {Measuring night-sky
  brightness with a wide-field CCD camera}.
\newblock \textit{Publications of the Astronomical Society of the Pacific},
  \textbf{119}, 192.

\bibitem[{Elvidge et~al.(2021)Elvidge, Zhizhin, Ghosh, Hsu and
  Taneja}]{elvidge2021annual}
Elvidge, C.~D., Zhizhin, M., Ghosh, T., Hsu, F.-C. and Taneja, J. (2021)
  {Annual time series of global VIIRS nighttime lights derived from monthly
  averages: 2012 to 2019}.
\newblock \textit{Remote Sensing}, \textbf{13}, 922.

\bibitem[{Finazzi(2020)}]{finazzi2020fulfilling}
Finazzi, F. (2020) Fulfilling the information need after an earthquake:
  Statistical modelling of citizen science seismic reports for predicting
  earthquake parameters in near realtime.
\newblock \textit{Journal of the Royal Statistical Society: Series A
  (Statistics in Society)}, \textbf{183}, 857--882.

\bibitem[{Fonte et~al.(2017)Fonte, Minghini, Patriarca, Antoniou, See and
  Skopeliti}]{fonte2017generating}
Fonte, C.~C., Minghini, M., Patriarca, J., Antoniou, V., See, L. and Skopeliti,
  A. (2017) {Generating up-to-date and detailed land use and land cover maps
  using OpenStreetMap and GlobeLand30}.
\newblock \textit{ISPRS International Journal of Geo-Information}, \textbf{6},
  125.

\bibitem[{Giraldo et~al.(2011)Giraldo, Delicado and
  Mateu}]{giraldo2011ordinary}
Giraldo, R., Delicado, P. and Mateu, J. (2011) Ordinary kriging for
  function-valued spatial data.
\newblock \textit{Environmental and ecological statistics}, \textbf{18},
  411--426.

\bibitem[{Gr{\"a}ler et~al.(2016)Gr{\"a}ler, Pebesma and
  Heuvelink}]{graler2016spatio}
Gr{\"a}ler, B., Pebesma, E.~J. and Heuvelink, G.~B. (2016) Spatio-temporal
  interpolation using gstat.
\newblock \textit{{The R Journal}}, \textbf{8}, 204--218.

\bibitem[{Hoek et~al.(2008)Hoek, Beelen, De~Hoogh, Vienneau, Gulliver, Fischer
  and Briggs}]{hoek2008review}
Hoek, G., Beelen, R., De~Hoogh, K., Vienneau, D., Gulliver, J., Fischer, P. and
  Briggs, D. (2008) A review of land-use regression models to assess spatial
  variation of outdoor air pollution.
\newblock \textit{Atmospheric environment}, \textbf{42}, 7561--7578.

\bibitem[{van Kesteren et~al.(2022)van Kesteren, Timmers, Garcia-Bernardo and
  Lugtig}]{erik_jan_van_kesteren_2022_6597859}
van Kesteren, E.-J., Timmers, A., Garcia-Bernardo, J. and Lugtig, P. (2022)
  {Code repository: correcting inferences for volunteer-collected data with
  geospatial sampling bias}.
\newblock \urlprefix\url{https://doi.org/10.5281/zenodo.6597859}.

\bibitem[{van Kesteren et~al.(2021)van Kesteren, Vida, de~Bruin and
  Oberski}]{van_kesteren_erik_jan_2021_4534188}
van Kesteren, E.-J., Vida, L., de~Bruin, J. and Oberski, D. (2021) {Enrich sf
  Data with Geographic Features from OpenStreetMaps}.
\newblock \urlprefix\url{https://doi.org/10.5281/zenodo.4534188}.

\bibitem[{Kuechly et~al.(2012)Kuechly, Kyba, Ruhtz, Lindemann, Wolter, Fischer
  and H{\"o}lker}]{kuechly2012aerial}
Kuechly, H.~U., Kyba, C.~C., Ruhtz, T., Lindemann, C., Wolter, C., Fischer, J.
  and H{\"o}lker, F. (2012) Aerial survey and spatial analysis of sources of
  light pollution in berlin, germany.
\newblock \textit{Remote Sensing of Environment}, \textbf{126}, 39--50.

\bibitem[{Land-Zandstra et~al.(2016)Land-Zandstra, Devilee, Snik, Buurmeijer
  and van~den Broek}]{land2016citizen}
Land-Zandstra, A.~M., Devilee, J.~L., Snik, F., Buurmeijer, F. and van~den
  Broek, J.~M. (2016) Citizen science on a smartphone: Participants’
  motivations and learning.
\newblock \textit{Public Understanding of Science}, \textbf{25}, 45--60.

\bibitem[{McKinley et~al.(2017)McKinley, Miller-Rushing, Ballard, Bonney,
  Brown, Cook-Patton, Evans, French, Parrish, Phillips
  et~al.}]{mckinley2017citizen}
McKinley, D.~C., Miller-Rushing, A.~J., Ballard, H.~L., Bonney, R., Brown, H.,
  Cook-Patton, S.~C., Evans, D.~M., French, R.~A., Parrish, J.~K., Phillips,
  T.~B. et~al. (2017) Citizen science can improve conservation science, natural
  resource management, and environmental protection.
\newblock \textit{Biological Conservation}, \textbf{208}, 15--28.

\bibitem[{Mooney and Minghini(2017)}]{mooney2017review}
Mooney, P. and Minghini, M. (2017) {A review of OpenStreetMap data}.
\newblock In \textit{{Mapping and the Citizen Sensor}}, 37--59. Ubiquity Press.

\bibitem[{Newman et~al.(2012)Newman, Wiggins, Crall, Graham, Newman and
  Crowston}]{newman2012future}
Newman, G., Wiggins, A., Crall, A., Graham, E., Newman, S. and Crowston, K.
  (2012) The future of citizen science: Emerging technologies and shifting
  paradigms.
\newblock \textit{Frontiers in Ecology and the Environment}, \textbf{10},
  298--304.

\bibitem[{{OpenStreetMap contributors}(2017)}]{Openstreetmap}
{OpenStreetMap contributors} (2017) {Planet dump retrieved from
  https://planet.osm.org }.
\newblock \url{ https://www.openstreetmap.org }.

\bibitem[{Petersen et~al.(2021)Petersen, Speed, Gr{\o}tan and
  Austrheim}]{petersen2021species}
Petersen, T.~K., Speed, J.~D., Gr{\o}tan, V. and Austrheim, G. (2021) Species
  data for understanding biodiversity dynamics: The what, where and when of
  species occurrence data collection.
\newblock \textit{Ecological Solutions and Evidence}, \textbf{2}, e12048.

\bibitem[{{R Core Team}(2021)}]{R}
{R Core Team} (2021) \textit{R: A Language and Environment for Statistical
  Computing}.
\newblock R Foundation for Statistical Computing, Vienna, Austria.
\newblock \urlprefix\url{https://www.R-project.org/}.

\bibitem[{Rambonnet et~al.(2019)Rambonnet, Vink, Land-Zandstra and
  Bosker}]{rambonnet2019making}
Rambonnet, L., Vink, S.~C., Land-Zandstra, A.~M. and Bosker, T. (2019) Making
  citizen science count: Best practices and challenges of citizen science
  projects on plastics in aquatic environments.
\newblock \textit{Marine pollution bulletin}, \textbf{145}, 271--277.

\bibitem[{de~Sherbinin et~al.(2021)de~Sherbinin, Bowser, Chuang, Cooper,
  Danielsen, Edmunds, Elias, Faustman, Hultquist, Mondardini
  et~al.}]{de2021critical}
de~Sherbinin, A., Bowser, A., Chuang, T.-R., Cooper, C., Danielsen, F.,
  Edmunds, R., Elias, P., Faustman, E., Hultquist, C., Mondardini, R. et~al.
  (2021) The critical importance of citizen science data.
\newblock \textit{Frontiers in Climate}, \textbf{3}, 20.

\bibitem[{Sonnenschein et~al.(2022)Sonnenschein, Scheider, de~Wit, Tonne and
  Vermeulen}]{sonnenschein2022agent}
Sonnenschein, T., Scheider, S., de~Wit, G.~A., Tonne, C.~C. and Vermeulen, R.
  (2022) Agent-based modeling of urban exposome interventions: prospects, model
  architectures, and methodological challenges.
\newblock \textit{Exposome}, \textbf{2}, osac009.

\bibitem[{Swanson et~al.(2016)Swanson, Kosmala, Lintott and
  Packer}]{swanson2016generalized}
Swanson, A., Kosmala, M., Lintott, C. and Packer, C. (2016) A generalized
  approach for producing, quantifying, and validating citizen science data from
  wildlife images.
\newblock \textit{Conservation Biology}, \textbf{30}, 520--531.

\bibitem[{Thieurmel and Elmarhraoui(2022)}]{thieurmel2022suncalc}
Thieurmel, B. and Elmarhraoui, A. (2022) \textit{suncalc: Compute Sun Position,
  Sunlight Phases, Moon Position and Lunar Phase}.
\newblock \urlprefix\url{https://CRAN.R-project.org/package=suncalc}.
\newblock R package version 0.5.1.

\bibitem[{Timmers and Lugtig(2021)}]{timmers_annemarie_2021_4724570}
Timmers, A. and Lugtig, P. (2021) {List of Citizen Science Projects in the
  Netherlands}.
\newblock \urlprefix\url{https://doi.org/10.5281/zenodo.4724570}.

\bibitem[{van Vliet and Moore(2016)}]{van2016citizen}
van Vliet, K. and Moore, C. (2016) Citizen science initiatives: Engaging the
  public and demystifying science.
\newblock \textit{Journal of microbiology \& biology education}, \textbf{17},
  13--16.

\bibitem[{Walker(1977)}]{walker1977effects}
Walker, M.~F. (1977) The effects of urban lighting on the brightness of the
  night sky.
\newblock \textit{Publications of the Astronomical Society of the Pacific},
  \textbf{89}, 405.

\bibitem[{Wickham et~al.(2014)Wickham, Homer, Vogelmann, McKerrow, Mueller,
  Herold and Coulston}]{wickham2014multi}
Wickham, J., Homer, C., Vogelmann, J., McKerrow, A., Mueller, R., Herold, N.
  and Coulston, J. (2014) {The multi-resolution land characteristics (MRLC)
  consortium—20 years of development and integration of USA national land
  cover data}.
\newblock \textit{Remote Sensing}, \textbf{6}, 7424--7441.

\bibitem[{Wu et~al.(2018)Wu, Zeng and Lung}]{wu2018hybrid}
Wu, C.-D., Zeng, Y.-T. and Lung, S.-C.~C. (2018) A hybrid kriging/land-use
  regression model to assess pm2. 5 spatial-temporal variability.
\newblock \textit{Science of the Total Environment}, \textbf{645}, 1456--1464.

\bibitem[{Yang et~al.(2018)Yang, Jin, Danielson, Homer, Gass, Bender, Case,
  Costello, Dewitz, Fry et~al.}]{yang2018new}
Yang, L., Jin, S., Danielson, P., Homer, C., Gass, L., Bender, S.~M., Case, A.,
  Costello, C., Dewitz, J., Fry, J. et~al. (2018) {A new generation of the
  United States National Land Cover Database: Requirements, research
  priorities, design, and implementation strategies}.
\newblock \textit{ISPRS journal of photogrammetry and remote sensing},
  \textbf{146}, 108--123.

\end{thebibliography}
\newpage
\appendix
\appendixpage

\section{Radiance}
\label{app:radiance}
\begin{figure}[H]
    \centering
    \includegraphics[width=\textwidth]{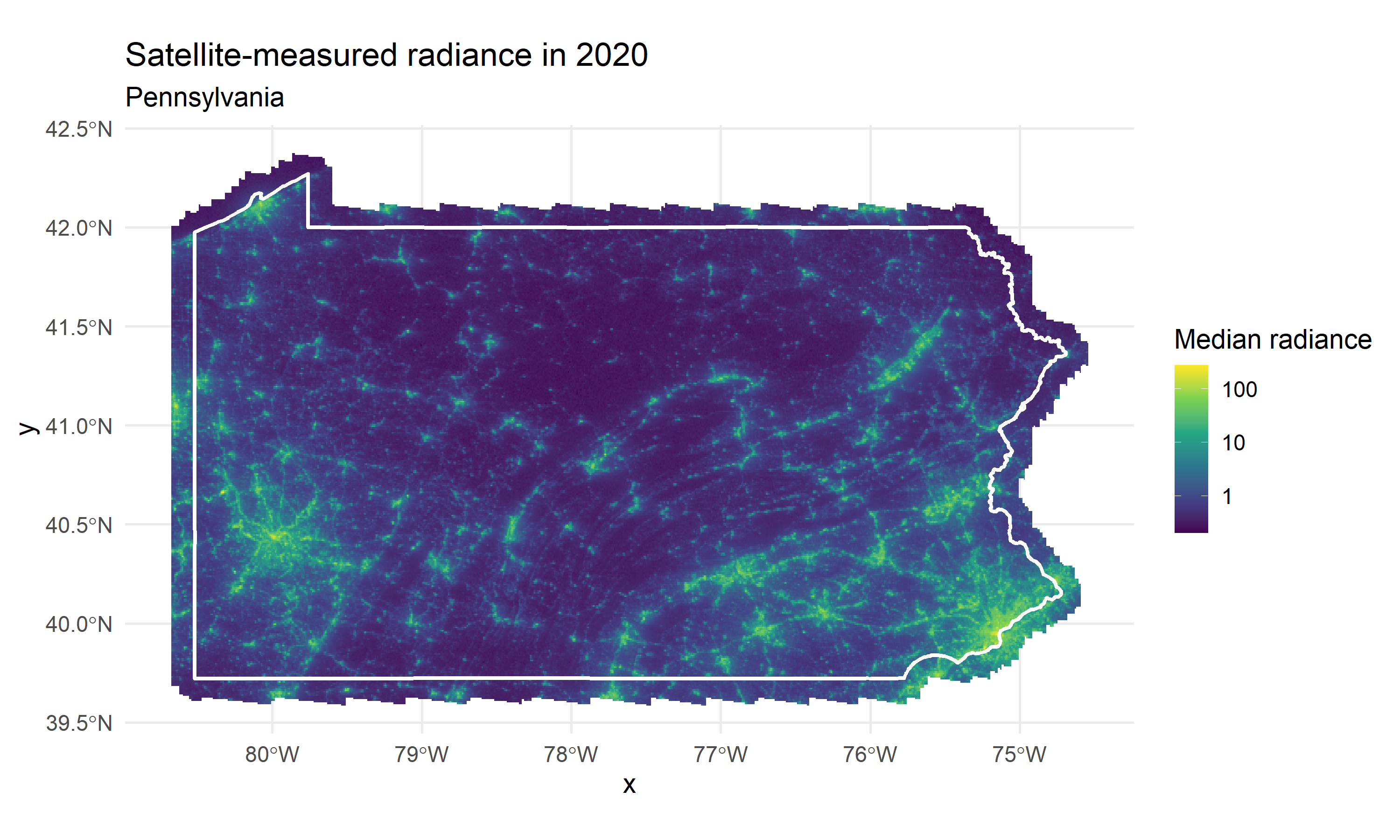}
    \caption{Raw radiance plot for Median nighttime cloud-free radiance in Pennsylvania based on the VIIRS imaging suite.}
    \label{fig:radiance}
\end{figure}

\section{Land cover types}
\label{app:landcover}
\begin{landscape}


\begin{table}
    \caption{\label{tab:landuse_info}Types of land-cover included in the United States National Land Cover Database.}
    \centering
    \large
    \begin{adjustbox}{width=\textwidth}
    \begin{tabular}{p{0.2\linewidth}p{0.9\linewidth}}
      \hline
      Label & Description \\ 
      \hline
      Open Water & areas of open water, generally with less than 25\% cover of vegetation or soil. \\ 
      Perennial Ice/Snow & areas characterized by a perennial cover of ice and/or snow, generally greater than 25\% of total cover. \\ 
      Developed, Open Space & areas with a mixture of some constructed materials, but mostly vegetation in the form of lawn grasses. Impervious surfaces account for less than 20\% of total cover. These areas most commonly include large-lot single-family housing units, parks, golf courses, and vegetation planted in developed settings for recreation, erosion control, or aesthetic purposes. \\ 
      Developed, Low Intensity & areas with a mixture of constructed materials and vegetation. Impervious surfaces account for 20\% to 49\% percent of total cover. These areas most commonly include single-family housing units. \\ 
      Developed, Medium Intensity & areas with a mixture of constructed materials and vegetation. Impervious surfaces account for 50\% to 79\% of the total cover. These areas most commonly include single-family housing units. \\ 
      Developed High Intensity & highly developed areas where people reside or work in high numbers. Examples include apartment complexes, row houses and commercial/industrial. Impervious surfaces account for 80\% to 100\% of the total cover. \\ 
      Barren Land (Rock/Sand/Clay) & areas of bedrock, desert pavement, scarps, talus, slides, volcanic material, glacial debris, sand dunes, strip mines, gravel pits and other accumulations of earthen material. Generally, vegetation accounts for less than 15\% of total cover. \\ 
      Deciduous Forest & areas dominated by trees generally greater than 5 meters tall, and greater than 20\% of total vegetation cover. More than 75\% of the tree species shed foliage simultaneously in response to seasonal change. \\ 
      Evergreen Forest & areas dominated by trees generally greater than 5 meters tall, and greater than 20\% of total vegetation cover. More than 75\% of the tree species maintain their leaves all year. Canopy is never without green foliage. \\ 
      Mixed Forest & areas dominated by trees generally greater than 5 meters tall, and greater than 20\% of total vegetation cover. Neither deciduous nor evergreen species are greater than 75\% of total tree cover. \\ 
      Dwarf Scrub & Alaska only areas dominated by shrubs less than 20 centimeters tall with shrub canopy typically greater than 20\% of total vegetation. This type is often co-associated with grasses, sedges, herbs, and non-vascular vegetation. \\ 
      Shrub/Scrub & areas dominated by shrubs; less than 5 meters tall with shrub canopy typically greater than 20\% of total vegetation. This class includes true shrubs, young trees in an early successional stage or trees stunted from environmental conditions. \\ 
      Grassland/Herbaceous & areas dominated by gramanoid or herbaceous vegetation, generally greater than 80\% of total vegetation. These areas are not subject to intensive management such as tilling, but can be utilized for grazing. \\ 
      Sedge/Herbaceous & Alaska only areas dominated by sedges and forbs, generally greater than 80\% of total vegetation. This type can occur with significant other grasses or other grass like plants, and includes sedge tundra, and sedge tussock tundra. \\ 
      Lichens & Alaska only areas dominated by fruticose or foliose lichens generally greater than 80\% of total vegetation. \\ 
      Moss & Alaska only areas dominated by mosses, generally greater than 80\% of total vegetation. \\ 
      Pasture/Hay & areas of grasses, legumes, or grass-legume mixtures planted for livestock grazing or the production of seed or hay crops, typically on a perennial cycle. Pasture/hay vegetation accounts for greater than 20\% of total vegetation. \\
      Cultivated Crops & areas used for the production of annual crops, such as corn, soybeans, vegetables, tobacco, and cotton, and also perennial woody crops such as orchards and vineyards. Crop vegetation accounts for greater than 20\% of total vegetation. This class also includes all land being actively tilled. \\ 
      Woody Wetlands & areas where forest or shrubland vegetation accounts for greater than 20\% of vegetative cover and the soil or substrate is periodically saturated with or covered with water. \\ 
      Emergent Herbaceous Wetlands & Areas where perennial herbaceous vegetation accounts for greater than 80\% of vegetative cover and the soil or substrate is periodically saturated with or covered with water. \\ 
      \hline
    \end{tabular}
    \end{adjustbox}
\end{table}
\end{landscape}

\begin{figure}
     \includegraphics[width=0.9\textwidth]{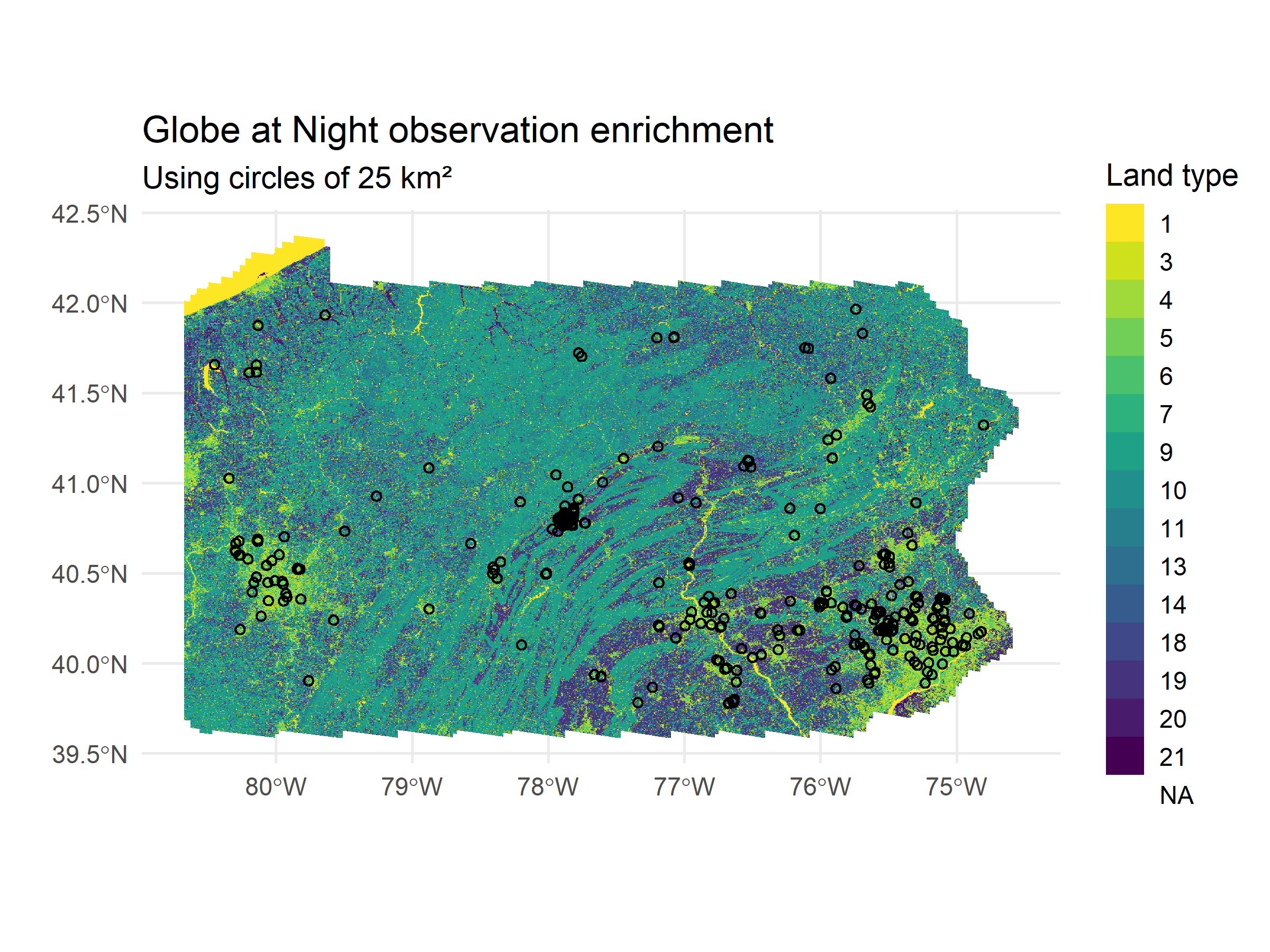}
    \caption{Enrichment procedure for the land-use data. For each observation we code the proportions in a radius covering 25 square kilometers}
    \label{fig:landuse_GaN}
\end{figure}
\begin{figure}
     \includegraphics[width=0.9\textwidth]{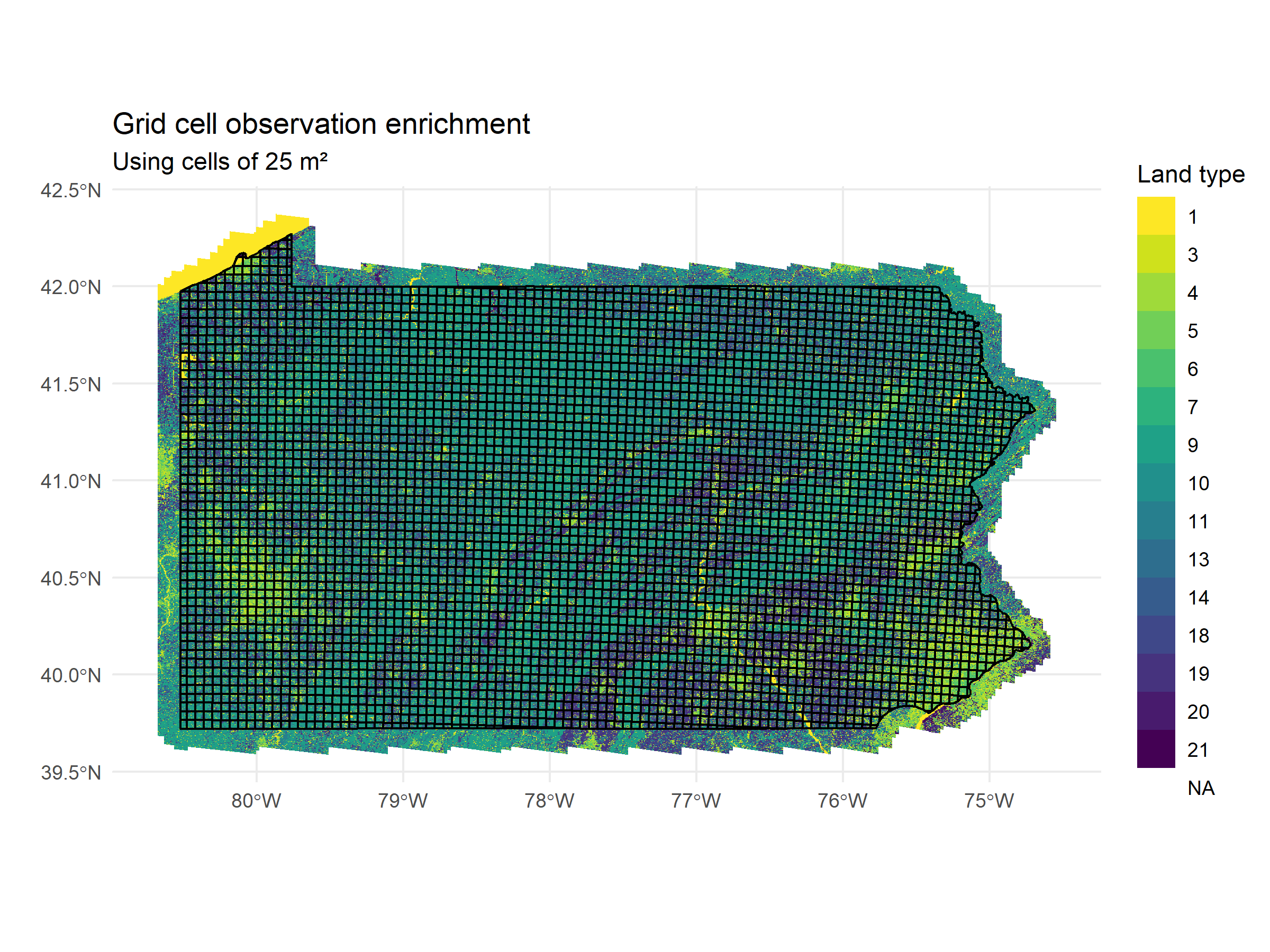}
    \caption{Enrichment procedure for the land-use data. For each cell in the raster covering the state of Pennsylvania, we measure land-use proportions.}
    \label{fig:landuse_state}
\end{figure}

\end{document}